\documentclass[twocolumn,amsmath,amssymb,nofootinbib,10pt]{revtex4-1}
\pdfoutput=1
\usepackage{graphicx,float,amsmath,amsmath, amssymb}
\usepackage[braket]{qcircuit}
\usepackage[export]{adjustbox}
\usepackage[breaklinks, colorlinks,linkcolor=black, citecolor=black,urlcolor = black]{hyperref}
\usepackage[labelsep=period]{caption}
\usepackage{subcaption}
\usepackage{wasysym}
\usepackage{multirow}
\usepackage{soul}

\graphicspath{{pictures/}}
\begin{document}
\title{Secure quantum communication through a wormhole}
	
\author{Grzegorz Czelusta}%
\author{Jakub Mielczarek}%
\email{jakub.mielczarek@uj.edu.pl}
\affiliation{Institute of Theoretical Physics, Jagiellonian University, {\L}ojasiewicza 11, 30-348 Cracow, Poland}
\date{\today}
\begin{abstract} 
An accumulation of theoretical evidence contribute to the picture of 
gravity as a manifestation of quantum entanglement in a certain 
many-body quantum system. This is in particular expresses in the 
ER=EPR conjecture, which relates gravitational Einstein-Rosen (ER) 
bridge with the Einstein-Podolsky-Rosen (EPR) quantum entangled 
pairs or, more generally, with the so-called Thermofield Double State.  
In this letter, the ER=EPR conjecture is employed to introduce unitary 
quantum teleportation protocol, which \emph{recycles} the entanglement 
forming traversable generalization of the Einstein-Rosen bridge. 
In consequence, the wormhole remains unaffected by the quantum 
teleportation. Furthermore, it is shown that the protocol guarantees the 
unconditional security of the quantum communication. Performance 
of the protocol is demonstrated in a simple setting with the use of 5-qubit 
Santiago IBM quantum computer, giving fidelities above the $2/3$ 
the classical limit for a representative set of teleported states. Security 
of the protocol has been supported by experimental studies performed 
with the use of the noisy quantum processor. Possible 
generalization of the protocol, which may have relevance in the 
context of macroscopic gravitational configurations, is also considered.  
\end{abstract}
\maketitle

\section{Introduction}

It is well known that quantum entanglement can be utilized to provide
an unconditionally secure method of communication. This is, in particular, 
the case in the entanglement-based quantum key distribution (QKD) 
\cite{Ekert1991}, complemented with the classical one-time pad (OTP) cipher. 
The same concerns the superdense coding \cite{Bennett1992}, which allows 
for secure exchange of classical bits, employing quantum entanglement.  
Moreover, thanks to the quantum teleportation protocol \cite{Bennett1993}, 
secure communication can be extended from bits to qubits. 

The problem with the above schemes is that a constant source of maximally 
entangled pairs is required for practical purposes. This is because the quantum 
entanglement is destroyed for every exchanged bit or qubit in a measurement 
process. This raises a quest to look for new secure quantum communication protocols,
which can \emph{recycle} the quantum entanglement and minimize the number 
of created entangled pairs.  A proposal for such a scheme in the case of the 
so-called port-based teleportation has been presented in Ref. \cite{Strelchuk2013}. 

Here, we introduce another entanglement-recycling quantum communication 
protocol, inspired by the ER=EPR conjecture \cite{Maldacena:2013xja}. The 
conjecture contributes to the duality between gravity and quantum entanglement, 
which has arisen from the studies AdS/CFT correspondence 
\cite{Maldacena:1997re} and holographic principle \cite{Susskind:1994vu}.  
In the ER=EPR conjecture, the Einstein-Podolsky-Rosen (EPR)
pairs are considered as being equivalent to the Einstein-Rosen 
(ER) bridge. The conjecture finds deep support in the case of the 
so-called anti-de Sitter (AdS) black hole \cite{Maldacena:2001kr}, 
which are, in fact, two back holes connected by the Einstein-Rosen 
bridge. It has been justified that the AdS black hole corresponds 
to the Thermofield Double State (TFD), which in the large temperature 
limit reduces to the product of the EPR pairs \cite{VanRaamsdonk:2010pw}.  
Further support to the ER=EPR comes, among the others, from 
the geometric interpretation of the entanglement entropy  \cite{Ryu:2006bv} 
and the relation between tensor networks and hyperbolic geometry \cite{Swingle:2009bg}. 
 
Following the ER=EPR conjecture, we will consider the TFD 
state of the AdS black hole. Then, quantum teleportation between two 
asymptotic regions connected by the ER bridge will be considered. 
In the usual case, such teleportation is associated with the intermediate 
measurement process, which leads to the emission of thermal Hawking radiation. 
This is, in particular, the case in the Hayden-Preskill protocol
\cite{Hayden:2007cs} in which measurement on a photon of 
Hawking radiation is necessary to accomplish the teleportation
\cite{Yoshida:2017non}. In contrast, in the case considered here, 
no measurement on the qubit of Hawking radiation is performed, 
so the process remains purely unitary. Thanks to this, the protocol can 
be designed such that the TFD state and the associated ER bridge 
are preserved during the teleportation.  Therefore, if a qubit is put 
into the black hole on one side of the bridge, it can be recovered 
on the other side without affecting the wormhole. 

Let us finally emphasize that at the classical level of General 
Relativity, the Einstein-Rosen, is considered as a non-traversable
wormhole, which is because of the Null Energy Condition. In 
consequence, a photon sent into such a wormhole from one 
side will never reach the other side. Therefore, from the perspective 
of causal communication through the wormhole, its traversable 
counterpart has to be considered. The necessary violation of 
the Null Energy Condition, while unrealistic classically, may 
occur when quantum gravitational effects are taken into account 
(see, e.g., \cite{Rubakov:2014jja,Horowitz:2019hgb}). In consequence, 
while referring to communication through the Einstein-Rosen bridge, 
we have in mind its traversable generalization. In the light of the 
gravity-entanglement duality, such traversable wormholes may 
play the role of quantum channels \cite{Susskind:2017nto,Bao:2018msr,Gao:2019nyj}. 

\section{The protocol}

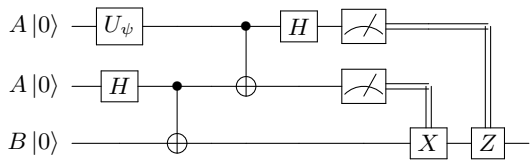
\begin{figure}[h]
	\leavevmode
	\centering
	\Qcircuit @C=1em @R=1em {
		\lstick{A \ket{0}} &\gate{U_\psi}  & \qw & \qw &\ctrl{1}&\gate{H} &\meter & \cw &\cw \cwx[2]\\
		\lstick{A \ket{0}} &\gate{H}&\ctrl{1}&\qw&\targ &\qw &\meter & \cw \cwx[1] \\
		\lstick{B \ket{0}} &\qw&\targ&\qw &\qw &\qw &\qw &\gate{X}&\gate{Z} &\qw  \\
	}
	\caption{The standard quantum teleportation protocol.}
	\label{fig:teleportation}
\end{figure}

Let us first remind the standard teleportation protocol, for which the 
corresponding quantum circuit is shown in Fig. \ref{fig:teleportation}. 
Here the operator $\hat{U}_\psi$ prepares the teleportated state, 
i.e. $|\psi \rangle = \hat{U}_\psi |0 \rangle$, belonging to let say Alice.
In order to teleport the state, Alice (A) and her friend Bob (B), 
share the EPR pair ($| \text{EPR} \rangle = \frac{1}{\sqrt{2}}\left(|00\rangle+|11\rangle\right)$),
which is then partially entangled with the state  $|\psi \rangle$. 
Subsequently, Alice performs measurements on her two qubits and sends 
the result to Bob via the classical channel. The associated two bits 
are used by Bob to accomplish the teleportation. 

\begin{figure}[ht!]
	\leavevmode
	\centering
	\Qcircuit @C=1em @R=1em {
		\lstick{}&\cw&\cw&\cw&\cw&\cw\cwx[2]&\\
		\lstick{}&\cw&\cw&\cw&\cw\cwx[1]&\\
		\lstick{\ket{0}} &\gate{H}&\ctrl{1}&\qw&\gate{X}&\gate{Z}&\ctrl{1}&\gate{H} & \qw \\
		\lstick{\ket{0}} &\qw&\targ&\qw&\qw&\qw&\targ&\qw & \qw \\
	}
	\caption{The superdense coding protocol.}
	\label{fig:superdensecoding}
\end{figure}
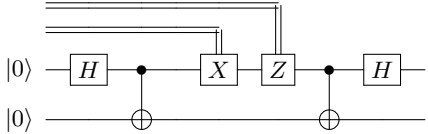

The first observation leading to our protocol is that the two classical 
bits can be encoded in a single qubit by virtue of the superdense coding.  
The protocol also utilizes the EPR pair, and the corresponding quantum 
circuit is shown in Fig. \ref{fig:superdensecoding}.

\begin{figure}[ht!]
	\leavevmode
	\centering
	\Qcircuit @C=0.8em @R=0.8em {
		\lstick{A \ket{0}} & \gate{U_\psi}& \qw& \qw& \qw & \qw&\ctrl{1}&\gate{H}& \ctrl{2}  & \qw \ar@{.}[]+<0.5em,1em>;[d]+<0.5em,-7em> &  \qw&   \qw & \qw  & \qw &  \qw &  &  \lstick{A} \\
		\lstick{A \ket{0}} & \gate{H} & \ctrl{3} & \qw&  \qw& \qw  & \targ & \ctrl{1}&  \qw&  \qw&  \qw&    \qw & \qw &  \qw & \qw &  &  \lstick{A} \\
		\lstick{A \ket{0}} & \gate{H} &  \qw& \ctrl{1} & \qw & \qw&  \qw& \targ&\ctrl{-2}&  \qw & \qw  &\ctrl{1}&\gate{H}& \ctrl{2}&  \qw &  &  \lstick{B}  \\
		\lstick{B \ket{0}} & \qw&  \qw& \targ & \qw&  \qw&  \qw &  \qw&  \qw& \qw& \qw & \targ & \ctrl{1}&  \qw&  \qw &  &  \lstick{B}  \\
		\lstick{B \ket{0}} & \qw& \targ &  \qw 	&  \qw & \qw &  \qw&  \qw&  \qw&   \qw&  \qw & \qw& \targ&\ctrl{-2}&  \qw &  &  \lstick{B}  \\
		\\
		\hspace{29.5em} \ket{\psi_{0}} 
		\gategroup{2}{2}{5}{4}{1.3em}{--}
		\gategroup{1}{7}{3}{9}{1.5em}{--}
		\gategroup{3}{12}{5}{14}{1.5em}{--}
	}
	\caption{The proposed protocol: Teleportation + Superdense coding = Teleportation without measurement. 
	Here $A$ denotes the qubits belonging to Alice and $B$ denotes the qubits belonging to Bob. Please notice
	that the middle qubit is exchanged between  Alice and Bob. The teleported state $|\psi \rangle = \hat{U}_\psi |0\rangle$ 
	is recovered in the final state bottom qubit.}
	\label{fig:ourteleportation}
\end{figure}
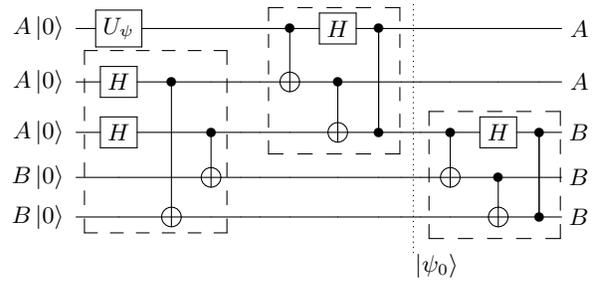

The second observation is that eventually, we can get rid of the measurement process and 
combine the two above protocols into the one shown in Fig. \ref{fig:ourteleportation}. 
The resulting protocol utilizes two EPR pairs to teleport one qubit using another qubit. 
Appropriate controlled gates have replaced the measurement operation.

In this scheme, Alice (sender) prepares the first qubit (the top one in Fig. \ref{fig:ourteleportation}) 
in the state  $|\psi \rangle$, which will be teleported. Furthermore, Alice shares two EPR pairs with Bob 
(receiver). Then, Alice applies a unitary operation to her three qubits and sends the third (from the top 
in Fig. \ref{fig:ourteleportation}) qubit to Bob. The unitary operation is given by the following action 
on the basis states: 
\begin{align}
&\hat{V}| a\rangle| b\rangle | c \rangle  = \widehat{\text{CZ}}_{13}(\hat{H}_1\otimes \widehat{\text{CNOT}}_{23})  
\widehat{\text{CNOT}}_{12}| a\rangle| b\rangle | c \rangle     \nonumber  \\
&=\frac{1}{\sqrt{2}}(-1)^{\bar{a} \cdot (a \oplus b \oplus c) }
| \bar{a} \rangle| a\oplus b \rangle | a \oplus b \oplus c \rangle \nonumber \\
&+\frac{1}{\sqrt{2}}(-1)^{a \cdot (a \oplus b \oplus c)+a}| a \rangle| a\oplus b \rangle | a \oplus b \oplus c \rangle,
\label{V}
\end{align}
where $a,b,c \in \{ 0,1 \}$. Here, $\hat{H}$ is the Hadamard operator,  $\widehat{\text{CNOT}}$ is the Controlled-NOT
operatora and $\widehat{\text{CZ}}$ stands for Controlled-Z operator and $\oplus$ stands for XOR operation. 

Finally, by applying the same $\hat{V}$ operation to his three qubits, Bob recovers his third qubit 
(the bottom qubit in  Fig. \ref{fig:ourteleportation}) in state $|\psi \rangle$. Moreover, the remaining 
qubits are two EPR pairs, as initially (up to SWAP between the first and the second qubit, to have 
the same order of pairs). In consequence, the EPR pairs can be reused to teleport another qubit. 
The action of the protocol associated with the circuit Fig. \ref{fig:ourteleportation} is, therefore, the 
following:
\begin{equation}
|\psi \rangle_A \otimes | \text{EPR} \rangle^{\otimes 2} \rightarrow |\text{EPR} \rangle^{\otimes 2}\otimes |\psi \rangle_B.
\end{equation} 

From the viewpoint of the security of the protocol, it has to be verified if any 
relevant information about the teleported state  $|\psi \rangle$ can be inferred 
from the qubit which Alice sends to Bob (the middle qubit in Fig. \ref{fig:ourteleportation}).
For this purpose one can trace the density matrix $\hat{\rho}_0 = |\psi _0 \rangle \langle \psi_0 |$ 
over all the qubits except the exchanged one. Here, $|\psi _0\rangle$ corresponds to the 
state between  Alice and Bob perform the $\hat{V}$ operation, as depicted in Fig. 
\ref{fig:ourteleportation}.

The resulting reduced density matrix for the exchanged qubit 
is $\hat{\rho}_{H} = \text{tr}_{1245} \hat{\rho}_0= \frac{1}{2} \mathbb{I}$ 
and the corresponding von Neumann entropy is $S\left(\hat{\rho}_{H}\right)
=-\text{tr}\hat{\rho}_{H}\ln\hat{\rho}_{H}=\ln 2$. The calculated entropy has 
maximal allowed value per single qubit. Therefore, the exchanged qubit is 
maximally entanglement with the rest of the system. In consequence, 
performing measurements on his qubit returns a random value of $0$ and $1$, 
with equal probability, independently on the details of the teleported state  
$|\psi \rangle$. This result will be supported by an experiment in the 
presence of noise, discussed in Sec. \ref{Experiment}. Furthermore, let 
us add that the exchanged qubit is an analog of the Hawking radiation photon, 
being in a thermal state. This justifies why the subscript $H$ has been used.

\section{Gravitational interpretation}

As we have already mentioned before, the introduced algorithm can be 
interpreted in terms of teleporting a quantum state through traversable 
wormhole. Let us now explore this connection. 

First of all, as discussed in Refs. \cite{Maldacena:2001kr,VanRaamsdonk:2010pw} 
the eternal AdS two sided black hole with wormhole is associated with the TFD state, 
which takes the following form:
\begin{equation}
|\text{TFD}\rangle :=\frac{1}{\sqrt{Z}} \sum_ne^{-\beta E_n}|E_n\rangle_A|E_n\rangle_B,
\label{TFD}
\end{equation}
where $\beta = \frac{1}{T}$ is an inverse of the temperature, $Z$ is the normalization 
factor and $E_n$ are energy eigenstates of the boundaries of the AdS black hole. The 
$|E_n\rangle_A$ and $|E_n\rangle_B$ are the energy eigenstates corresponding to
the sites of Alice and Bob respectively. The energy eigenstates  $|E_n\rangle$ are
considered to be the basis states of the quantum register of $N$ qubits. In consequence,
the index $n$ runs from $0$ to $2^N-1$. The spacial case of large temperature limit 
$(T \rightarrow \infty)$ will be of importance for our further discussion. In this case,
the state (\ref{TFD}) simplifies as follows:
\begin{align}
|\text{TFD}\rangle \xrightarrow[T\rightarrow\infty]{}\sum_n|E_n\rangle|E_n\rangle = |\text{EPR}\rangle^{\otimes N}.
\end{align}
For $N=2$, the hight-temperature $|\text{TFD}\rangle$ is just the state
we considered in the previous section to be formed between Alice and Bob.  In that 
case one can write $|E_0\rangle=|00\rangle$, $|E_1\rangle=|01\rangle$, $|E_2\rangle=|10\rangle$ and 
$|E_3\rangle=|11\rangle$. 

Then, the state $|\psi\rangle$ can be viewed as a state which falls into the black hole from 
one boundary, travels through the wormhole, and then reaches the second boundary. The 
qubit which is sent from Alice to Bob can be interpreted as a photon of 
Hawking radiation, which allows teleporting the state $|\psi \rangle $ through the bridge. 
Notably, in the considered protocol, no measurement on the Hawking photon is performed. 

The traced state representing one side of black hole, for example Alice part of EPR pairs, is 
maximally mixed: $\hat{\rho}_{A}=\text{tr}_B\hat{\rho}_{AB}=\frac{1}{2^N}\hat{\mathbb{I}}=\rho_{B}$, 
where $\hat{\rho}_{AB}:=|\text{TFD}\rangle\langle \text{TFD}|$.  Entropy of this state
is maximal and equal to $S\left(\rho_A\right)=-\text{tr}\rho_A\log\rho_A= N \ln 2$. Therefore,
entropy of each black hole is equal to the number of EPR pairs. Supposing any qubit, contributing 
to the pair, covers a fixed area, e.g. the Planck area, the entropy is proportional to total 
area of the black hole horizon, in agreement with the Bekenstein-Hawking formula.  

\section{Experimental demonstration}
\label{Experiment}

Thanks to the recent developments in quantum computing technologies, the quantum 
protocol introduced in the previous sections can already be studied
experimentally. For this purpose, we have employed superconducting Santiago 
5-qubit quantum computer provided by IBM. The quantum computer is characterized by 
relatively high \emph{quantum volume} equal 32 \cite{IBM}. The topology of the quantum 
processor is linear. Please notice that the related Hayden-Preskill protocol has already 
been simulated on a 7-qubit ion trap quantum computer \cite{Landsman:2018jpm}. 
Furthermore, quantum simulations of the TFD states have recently been discussed 
in Refs. \cite{Brown:2019hmk,Nezami:2021yaq,Blok:2020may}

In our studies, the quantum teleportation protocol (\ref{fig:ourteleportation}), 
has been performed for 6 states $\{ |0\rangle, |1\rangle, |+\rangle =
\frac{|0\rangle+|1\rangle}{\sqrt{2}}, |-\rangle = \frac{|0\rangle-|1\rangle}{\sqrt{2}}, 
|\leftturn\rangle =\frac{|0\rangle-i|1\rangle}{\sqrt{2}} ,  
|\rightturn\rangle =\frac{|0\rangle+i|1\rangle}{\sqrt{2}} \}.$
The states form a representative set over the Bloch sphere.  

\begin{figure}[ht!]
	\includegraphics[scale=0.35]{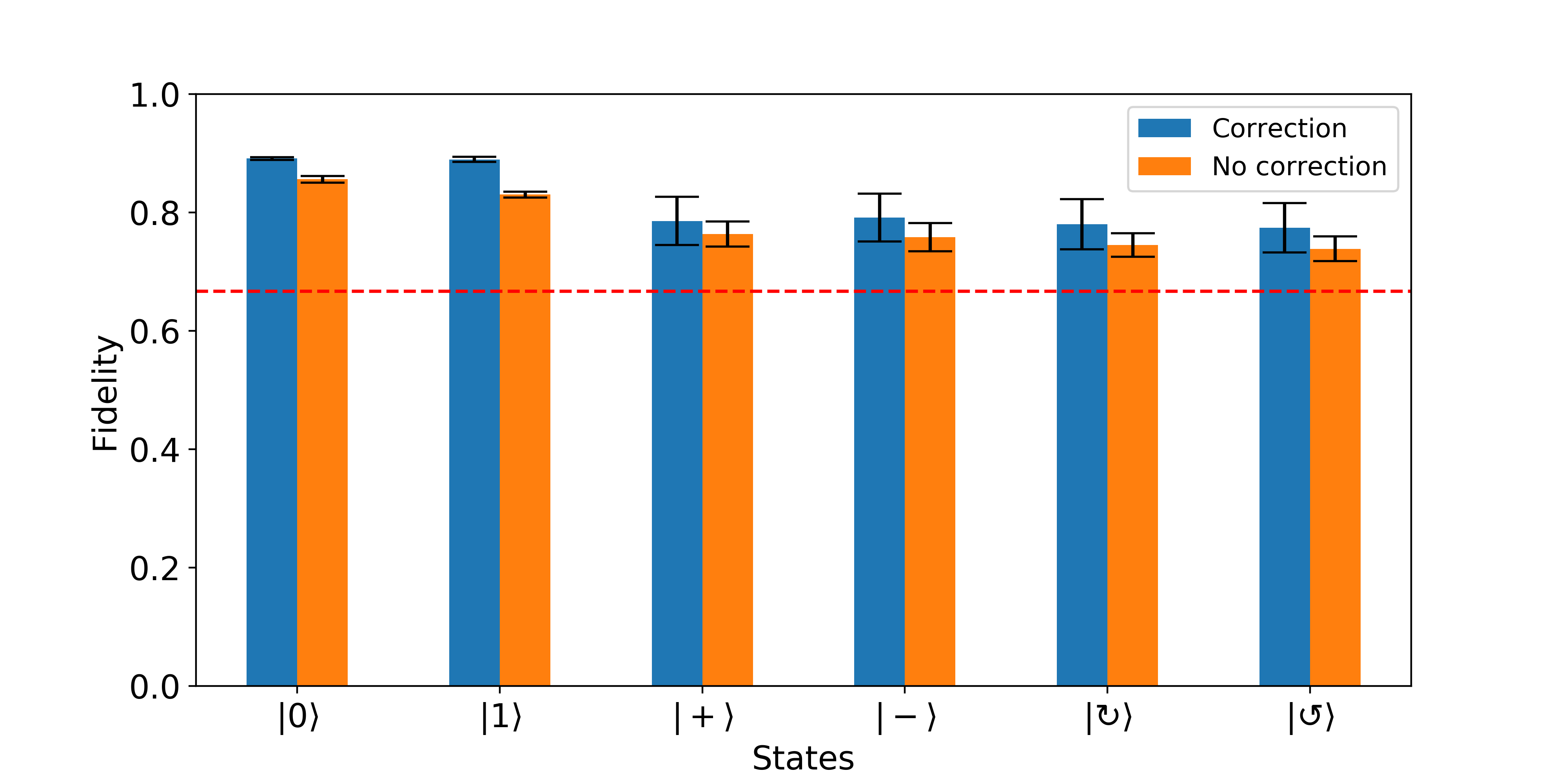}
	\caption{Quantum fidelities of the states teleported with the use of the protocol depicted in 
	Fig. \ref{fig:ourteleportation}. The results have been obtained with the use of 5-qubit 
	IBM Santiago superconducting processor. The dashed horizontal line corresponds to 
	the quantum limit $2/3$.}
	\label{TeleFidelities}
\end{figure}

The teleported states have been compared with the reference states employing the 
quantum fidelity measure:
\begin{equation}
F(\hat{\rho}_1,\hat{\rho}_2) :=\left(\text{tr}\sqrt{\sqrt{\hat{\rho}_1}\hat{\rho}_2\sqrt{\hat{\rho}_1}}\right)^2.
\label{fidelity}
\end{equation}
The experimental density matrix has been recovered by performing quantum 
tomography on the teleported qubit (the bottom qubit in  Fig. \ref{fig:ourteleportation}).
For each of the teleported states 12 runs (each with 8192 shots) of the algorithms have 
been made. Furthermore, both the case with and without applying measurement 
error mitigation were considered \cite{Mitigation}.  The resulting fidelities are shown 
in Fig. \ref{TeleFidelities}.

In all of the considered cases, the obtained fidelities exceed the optimal 
classical strategy, which is $F=2/3$ \cite{Massar}.

Furthermore, in order to experimentally verify security of the protocol in 
presence of errors, measurement on exchanged Hawking qubit were 
performed. Performing 12 runs (each with 8192 shots) for each of the teleported 
states, the obtained fidelities of the Hawking qubit, with respect to 
the density matrix $\hat{\rho}_{H} = \frac{1}{2} \mathbb{I}$, were collected 
in Tab. \ref{Table1}.

\begin{table}
	\begin{tabular}{|c|c|c|}
		\hline
		State & Fidelity & Entropy$/\ln 2$   \\
		\hline
		$|0\rangle$ & $0.99969\pm0.00022$ &  $0.99911\pm0.00061$ \\
		$|1\rangle$ & $0.99950\pm0.00022$ & $0.99854\pm0.00063$ \\
		$|+\rangle$ & $0.99978\pm0.00010$ & $0.99937\pm0.00030$ \\
		$|-\rangle$ & $0.99977\pm0.00015$ & $0.99933\pm0.00043$ \\
		$|\circlearrowright\rangle$ & $0.99911\pm0.00030$ & $0.99745\pm0.00084$ \\
		$|\circlearrowleft\rangle$ & $0.99944\pm0.00021$ & $0.9984\pm0.0006$ \\
		\hline
	\end{tabular}
	\caption{Uncorrected quantum fidelities and entropies (divided by $\ln2$) of 
	the intermediate (Hawking) qubit. The results were obtained on the 5-qubit 
	IBM Santiago quantum processor.}
\label{Table1}
\end{table}

The obtained values are equal $1$ with accuracy at the level of the order of $10^{-3}$. 
Furthermore, the dispersion of the mean values of the measured fidelities is approximately 
equal to $0.0002$. Therefore, taking into account the experimental errors, no worrying 
differences between the results for different teleported states have been observed. 

Similar conclusions concern entropies of the Hawking qubit, values of which 
are also collected in Tab. \ref{Table1}. The fact that the values of entropies are 
not affected by noise in the system is supported by the pseudo-pure state 
approximation of the thermal noise. In this case, the mixed-state density 
matrix corresponding to the pure state $|\psi_0\rangle$ is 
$\hat{\rho}_{\epsilon} := (1-\epsilon) \hat{\rho}_{0} + \frac{\epsilon}{2^5} \hat{\mathbb{I}}$,   
where $\epsilon \in [0,1]$ parametrizes departure from the pure state.
Because of linearity of the trace, the reduced density matrix for the 
Hawking qubit (the 3rd in Fig. \ref{fig:ourteleportation}) is: 
\begin{equation}
\text{tr}_{1245} \hat{\rho}_{\epsilon} = (1-\epsilon) \text{tr}_{1245}\hat{\rho}_{0} 
+ \frac{\epsilon}{2} \hat{\mathbb{I}} = \frac{1}{2} \hat{\mathbb{I}},   
\end{equation}
where we used the already discussed fact that $\text{tr}_{1245}\hat{\rho}_{0} = \frac{1}{2} \hat{\mathbb{I}}$.
In consequence, the entropy of the exchanged (Hawking) qubit remains unaffected 
by the thermal noise in the pseudo-pure state approximation. 

\section{Possible generalization}

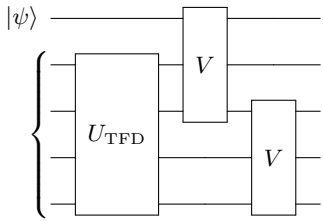
\begin{figure}[ht!]
	\leavevmode
	\centering
	\Qcircuit @C=1em @R=1em {
		\lstick{\ket{\psi}} &\qw &\multigate{2}{V}&\qw&\qw\\
		&\multigate{3}{U_{\text{TFD}}}&\ghost{V}&\qw&\qw\\
		 &\ghost{U_{\text{TFD}}}&\ghost{V}&\multigate{2}{V}&\qw\\
		 &\ghost{U_{\text{TFD}}}&\qw&\ghost{V}&\qw\\
		&\ghost{U_{\text{TFD}}}&\qw&\ghost{V}&\qw 
		\gategroup{2}{1}{5}{1}{1em}{\{}
		}
	\caption{General form of the considered protocol.}
	\label{fig:generalprotocol}
\end{figure}

One can think about generalizing our protocol to cases with bigger black holes, i.e., 
with more EPR pairs. We aim to find a unitary operator $\hat{V}$, which extends the 
one given by  Eq. \ref{V} to a higher number of qubits. The generalized protocol 
can be then related to the quantum circuit presented in Fig. \ref{fig:generalprotocol}.

From the perspective of the black hole physics, the operator $\hat{V}$ is expected to be the so-called 
scrambling operator of a horizon \cite{Sekino:2008he}. However, analysis of the related complexity of 
the operator goes beyond the scope of this letter and will be addressed elsewhere. Here, our main 
focus is on verifying whether the operator $\hat{V}$ for a higher number of qubits can be found at all. 

Our strategy is to determine the form of $\hat{V}$ using variational methods, which can be applied on a 
quantum computer or on a classical simulator of a quantum computer. We prepare quantum circuits 
consist of $2N+1$ qubits, which has first qubit in state $|\psi\rangle$ and $2N$ qubits in the $|\text{TFD}\rangle$  
state. Then, we apply a parametrization ansatz for $\hat{V}$ on the first $N+1$ qubits. Finally, the same 
ansatz with the same parameters is applied on the last $N+1$ qubits. 

When our ansatz with given parameters is a good approximation of $\hat{V}$, the final state 
is expected to be $|\psi_{\text{final}} \rangle := | \text{EPR} \rangle^{\otimes N} \otimes | \psi  \rangle$. 
We require that our protocol works for any qubit state $|\psi \rangle$, so we have to check 
fidelity for a set of $n$ states. Furthermore, for security reason, we also want to have the 
intermediate qubit being in maximally entanglement with the wormhole, which results in additional 
entropy term to the cost function. The proposed cost function is:
\begin{equation}
	C=1-\frac{1}{2}\frac{\sum_{k=1}^n\left(F\left(\hat{\rho}_{\text{final}},\hat{\rho} \right)+
	S\left(\hat{\rho}_H\right)/\ln 2\right)}{n},
	\label{CostFunction}
\end{equation}
where the $1/2$ factor ensures that the function is non-negative and is equal zero for the optimal 
case. Here $\hat{\rho}$ is the density matrix of the measured state and $\hat{\rho}_{\text{final}} 
= |\psi_{\text{final}} \rangle \langle \psi_{\text{final}} |$ is the desired final density matrix. 

\begin{figure}[ht!]
\leavevmode
\centering
\Qcircuit @C=0.5em @R=1.0em {
	\lstick{ {q}_{0}} & \gate{{\ensuremath{\theta}}_0} & \ctrl{1} & \gate{{\ensuremath{\theta}}_4} & \qw & \ctrl{1} & \gate{{\ensuremath{\theta}}_8} & \qw & \qw & \qw & \qw\\
	\lstick{ {q}_{1}} & \gate{{\ensuremath{\theta}}_1} & \targ & \ctrl{1} & \gate{{\ensuremath{\theta}}_5} & \targ & \ctrl{1} & \gate{{\ensuremath{\theta}}_9} & \qw & \qw & \qw\\
	\lstick{ {q}_{2}} & \gate{{\ensuremath{\theta}}_2} & \qw & \targ & \ctrl{1} & \gate{{\ensuremath{\theta}}_6} & \targ & \ctrl{1} & \gate{{\ensuremath{\theta}}_{10}} & \qw & \qw\\
	\lstick{ {q}_{3}} & \gate{{\ensuremath{\theta}}_3} & \qw & \qw & \targ & \gate{{\ensuremath{\theta}}_7} & \qw & \targ & \gate{{\ensuremath{\theta}}_{11}} & \qw & \qw\\
}
\caption{The ansatz $R_Y$ for four qubits with 2 repetitions. The boxes are $R_y({\ensuremath{\theta}})$ rotation gates.}
\label{RYansatz}
\end{figure}
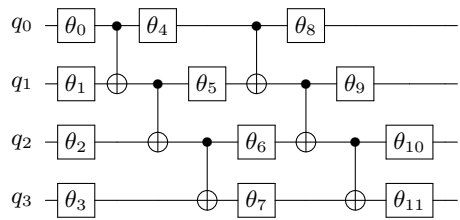

In what follows, we show results of the procedure for the case of $N=3$, which is a first step 
beyond the case considered so-far.  Furthermore, the $n=6$ states considered in Sec. \ref{Experiment} 
are used in the averaging procedure. Concerning the assumed form of the $\hat{V}$ operator 
the so-called $R_Y$ ansatz is considered, for which the quantum circuit is shown in Fig. \ref{RYansatz}. 

\begin{figure}[ht!]
\includegraphics[scale=0.35]{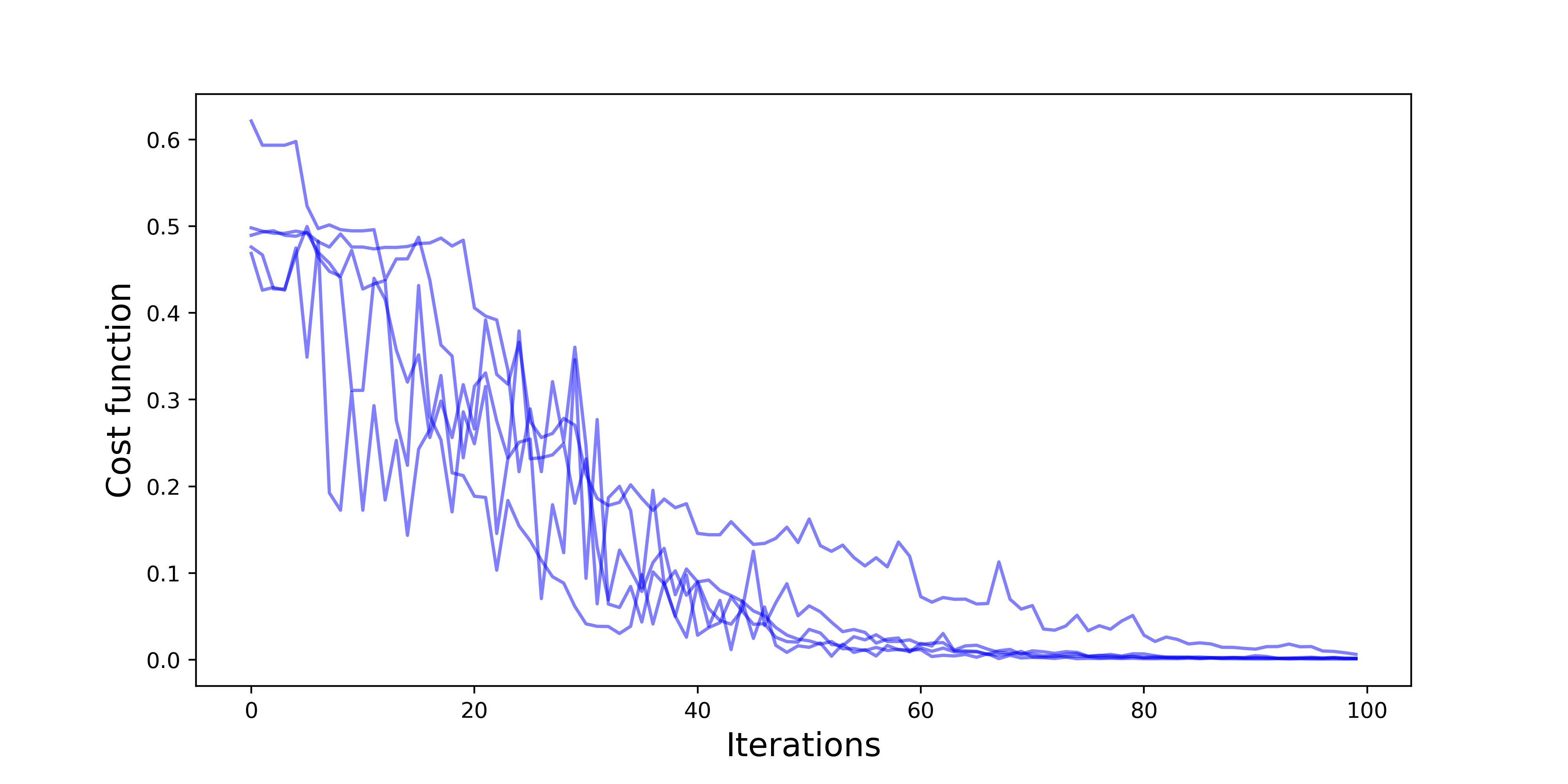}
\caption{Plot of the cost function (\ref{CostFunction}) for the initial 100 optimization steps for the five 
representative runs.}
\label{OptimizationPlot}
\end{figure}

In this case, the parameter space is given by the 12 angles $\{ \theta_0,\theta_1,..., \theta_{11} \}$,
which significantly reduces the number of $30$ real parameters for a general case (excluding the 
total phase and normalization). The initial values are randomly choosen from flat distributions for 
each of the angles, $\theta \in [0, 2\pi)$. Evolution of the cost function (\ref{CostFunction}) for 
exemplary five optimization runs are shown in Fig. \ref{OptimizationPlot}

In total, 500 optimization steps have been performed with the use of the Constrained 
Optimization by Linear Approximation (COBYLA) algorithm \cite{Powell}. The resulting 
values of the five cases are collected in Tab. \ref{Table2}. The results confirm that the 
protocol under consideration can be successfully extended beyond the $N=2$ case. Worth 
emphasizing is that the extension is non-trivial, e.g., is not just a result of adding ``disconnected'' 
EPR pair to the protocol shown in Fig. \ref{fig:ourteleportation}.
  
\begin{table}[ht!]
	\begin{tabular}{|c|c|c|c|}
		\hline
		Run & Cost function & Fidelity & Entropy$/\ln 2$\\
		\hline
		1.&$2.8\cdot 10^{-5}$&$1$&$0.999945\pm 7.7\cdot 10^{-5}$\\
		2.&$2.9\cdot 10^{-5}$&$1$&$0.999942\pm 5.1\cdot 10^{-5}$\\
		3.&$5.0\cdot 10^{-5}$&$1$&$0.999903\pm 7.3\cdot 10^{-5}$\\
		4.&$2.8\cdot 10^{-5}$&$1$&$0.999944\pm 4.6\cdot 10^{-5}$\\
		5.&$3.1\cdot 10^{-5}$&$1$&$0.999938\pm 5.8\cdot 10^{-5}$\\		\hline
	\end{tabular}
	\caption{Values of fidelity and cost function for the final state and 
	entropy of the ``Hawking qubit''  obtained in 500 iterations of the 
	optimization. The optimization has been performed with the use 
	of simulator of a quantum computer.}
	\label{Table2}
\end{table}

One can show that the $R_Y$ ansatz for $N=2$ and with three repetitions covers 
the case of operator $\hat{V}$ defined in Eq. \ref{V}. As we have checked, the 
$N=2$ case cannot be recovered with two repetitions. However, for the higher 
than two number of EPR pairs, the  $R_Y$ ansatz is a promising candidate for 
the general form of the  $\hat{V}$ operator. Further studies are needed to investigate
properties of the general  $\hat{V}$ operator in details, including its possible 
action as a scrambling operator. 

\section{Summary}

In this letter, we have introduced an entanglement-recycling quantum teleportation protocol
inspired by the considerations of the AdS black hole. The protocol works in analogy to
sending a qubit through a traversable gravitational wormhole, which in our studies corresponds 
to the large temperature limit of the Thermofield Double State. The protocol has been tested on a 
representative set of states with the use of a 5-qubit IBM Santiago quantum computer, 
giving fidelities above the classical best strategy value. This confirms that the algorithm can 
already be implemented with the use of currently available quantum technologies. 
Furthermore, it has been justified both theoretically and by performing a quantum 
experiment that the protocol is secure with respect to performing measurements on the 
exchanged qubit. Finally, possible generalization of the protocol to the case with a higher 
number of EPR pairs has been investigated. The generalizations, which eventually may 
allow us to apply our results to the case of macroscopic AdS black hole, require further, 
detailed studies. Some of the remaining open theoretical issues are: answering whether 
the protocol can be generalized such that the $\hat{V}$ part acts as a scrambling operator, 
fate of the protocol in finite-temperature regime ($T< \infty$), and teleportation of higher 
dimensional quantum states (not only qubits). On the other hand, from the practical viewpoint,
 an open challenge is the photonic implementation of the introduced protocol.

\section*{Acknowledgements}

Authors are supported by the Sonata Bis Grant DEC-2017/26/E/ST2/00763 of the 
National Science Centre Poland. Furthermore, this publication was made possible 
through the support of the ID\# 61466 grant from the John Templeton Foundation, 
as part of the ``The Quantum Information Structure of Spacetime (QISS)'' Project 
(qiss.fr). The opinions expressed in this publication are those of the authors and 
do not necessarily reflect the views of the John Templeton Foundation.
This publication has also been supported by the SciMat Priority Research Area 
budget under the program ``Excellence Initiative - Research University'' at the 
Jagiellonian University.

\end{document}